# Bibliometric Analysis of NIME References and Citations

Stefano Fasciani
Department of Musicology
University of Oslo
Oslo, Norway
stefano.fasciani@imv.uio.no

## ABSTRACT
This paper presents a bibliometric analysis that examines the works cited in, as well as those citing, NIME papers; for brevity, we refer to these as 'references' and 'citations'. Utilizing existing tools, we have computationally extracted data from the NIME proceedings archive and retrieved metadata from an academic database, including details of associated references and citations. From this data, we computed a range of metrics and statistics, which we present in this paper. We offer quantitative insights into NIME as a scholarly publication venue, its connections to other venues, and its relationship with various fields of study and authors. Based on our data interpretations, we provide several recommendations for the community's future. In sharing the software we developed for this study, and the summarized raw data, we enable other NIME researchers to conduct more in-depth investigations and examine specific trends.

## Author Keywords
bibliometric study, proceedings analysis, scholar citations, meta study

## CCS Concepts
•Applied computing → Arts and humanities → Sound and music computing; •General and reference → Document types → Surveys and overviews;

## 1. INTRODUCTION
Since its beginning in 2001, more than two thousand papers have been published in the proceedings of the International Conference on New Interfaces for Musical Expression (NIME). In recent years, the NIME community has shown a strong inclination toward self-reflection, which has also led to several meta-studies and publications in this area. Masu et al. [1] have identified a general lack of environmental-related considerations in NIME research practices and proposed a framework to increase awareness and engagement with ecological and environmental issues. In a subsequent publication, Masu et al. [2] further developed this idea by advocating for NIME research to adopt a less singular focus on 'newness', suggesting that significant contributions to the field can also be achieved by reusing, updating, augmenting, and committing to long-term engagement with existing instruments. However, pursuing this line of research necessitates replication of existing designs, and systematic documentation is crucial. In this context, Calegario et al. [3] have noted that recent NIME publications still lack the necessary documentation to enable replicability.

Morreale et al. [4] argue that while the NIME is a mature research community, it has tended to be inward-looking and has rarely engaged with external trends. They call for the community to broaden its discussions to address the political dimensions involved in new musical instruments also made outside the NIME context. The limited engagement with broader technological and societal issues is a concern that the NIME community has yet to fully address.

Research on musical interfaces has a long history, with a significant amount of research material published before 2001. Wanderley [5] provided a comprehensive review of key early works and developments, as well as significant pre-NIME publications venues related to musical interfaces. He urges the NIME community to recognize this vast body of early research to avoid 're-inventing the wheel', to consolidate pre- and off NIME works in one place, to track the evolution of technical terms, to appreciate topics that may not be popular in mainstream publications, to access important works published in other languages, and to cultivate a more inclusive narrative about the origins of NIME.

Previously, we conducted a computational analysis of the proceedings of the first 20 editions of the NIME conference, providing a range of figures and metrics on papers, authorship, affiliations, travel, and topics, as well as their geographical and temporal distribution [6]. This work also includes statistics on the number of citations received by NIME papers, detailing the tally, distribution, and impact across the various editions of the conference.

In this paper, we extend our previous work [6] by conducting a more comprehensive and detailed analysis of the scholarly works cited in NIME proceedings, as well as those that cite NIME papers. For the sake of brevity, we will use the term 'reference' to denote a work that is cited in a NIME paper and thus appear in its reference list. Conversely, we will use 'citation' to refer to a scholarly work that cites a NIME paper. In particular, we begin by extracting metadata and unique identifiers for NIME references and citations from a public database. The metadata includes details such as authorship, field of study, publication year, publication venue, and embeddings. We then process, filter, cross-reference, and sort this data computationally to analyze various aspects: citation and reference distribution across NIME papers; the extent of proceedings self-citation; academic disciplines and publication venues linked to NIME through scholarly citations; and seminal works significant to the NIME community.

One of our objective and contribution is to offer the NIME community a collection of objective figures, beyond the summaries provided in this manuscript. Comprehensive tables and the software used for our computational analysis are readily available online[1], for purposes of replication or further extension of this analysis in the future. In this paper, we present figures that provide quantitative insights into NIME's role as a scholarly publication venue, as well as its connections to other venues, fields of study, and authors. We interpret data on references and citations to reveal both known and potentially unknown facts of the NIME scholarly community. It is important to note that these interpretations may sometimes lead to conflicting perspectives. Additionally, based on our analyses, we offer several recommendations for the community's future. This work also aims to serve as an objective foundation for the assertions made in recent works reflecting on the past and future of NIME [4], [5].

In Section 2, we present the methodology used to develop this study. Sections 3 and 4 summarize the data extracted for references and citations. A discussion on findings and and limitations of this work are presented in Section 5, including possible future extensions.

[1] https://github.com/jacksongoode/NIME-proceedings-analyzer

## 1.1 Scholarly references and citations

References are essential components in scientific literature and scholarly publications. Despite the differences in citation styles used across disciplines, referencing is commonly used to acknowledge other's work, to differentiate one's work from existing ones, or as a supporting argument [7]. Moreover, through references readers can access original sources and get informed about related theories, methods, data or results. Almost 50 years ago, Gilbert [8] argued that references also serve more subtle purposes: coping with intellectual property issues and increase the persuasiveness of a paper. Indeed, the quality of a paper is often judged not only by its content but also by the size of its reference list and the prestige of the publishers cited. During the review process, cross-checking the accuracy of citations and the references is complex, and often these details are not thoroughly verified. Errors in citations are common and can be perpetuated across academic papers, with studies suggesting that 70 to 90% of citations are copied from reference lists of other papers [9]. This practice of copying citations, combined with the existing favorable reputation of some publications, can significantly skew citation counts within a particular field or venue [7], [8], [9]. Indeed, the common assumption that citations in journals or conference proceedings follow a normal distribution is typically incorrect. Therefore, relying on impact factors to gauge the quality of individual articles is fundamentally flawed [10]. To date, various theories have been proposed to understand the nature of citations in scientific literature [11], [12], [13], [14]. These theories often treat science as a social system and require systematic analysis of reference-related data within specific academic fields or publication settings. It is not feasible to generalize findings because citation behaviors and statistics vary greatly across different disciplines [15].

References are the foundation upon which a paper is built, and these in turn become the basis for future works that cite it. Consequently, each paper serves as both a convergence point in the network of existing knowledge and a divergence point for the creation of new knowledge. By analyzing the works both cited in and citing a specific paper, it is possible to derive insights into how these works are interconnected and the indirect influence they may have on one another. When this analysis is extended to an entire publication venue, we can gain deeper insights into the impact of the referenced journals or conference proceedings.

# 2. METHODOLOGY

The corpus of papers published in the NIME proceedings is so large that manual data extraction is practically infeasible. Therefore, we employ computational methods to mine existing databases and analyze data related to references and citations. As starting point, we utilized our *NIME Proceedings Analyzer*[1] (NIME PA) [16], a suite of Python methods designed to aggregate, scrape and retrieve meta-data related to NIME papers directly from the publicly available list of community-compiled BibTeX entries[2]. The extracted metadata is organized into a tabular data structure, and the textual content of the papers is structured into a collection of text files. These resources are then analyzed to yield a range of bibliometric figures and statistics. Specifically for this study, the tabular data output from the previous version of the NIME PA included only: the total number of citations and the total number of highly influential citations (i.e., instances where the cited publication has a significant impact on the citing publication), as retrieved from Semantic Scholar[3].

Among the various academic search engines, Semantic Scholar was selected because it offers the most comprehensive and accurate indexing of NIME papers and provides an Application Programming Interface (API) [16].

In Semantic Scholar, papers and authors are each associated with a unique identifier, which allows for reliable mining of the database across the network of references and citations. However, accurately identifying the correct paper using the information in the BibTeX file, which primarily includes authors and title, is not a trivial task. Successful identification can be hindered by several issues: incorrect data entry in the BibTeX file, errors in how the paper is registered in Semantic Scholar, and inconsistent handling of non-ASCII characters in titles and authors' names. Works with a single author and a short title pose distinct challenges, as they often lead to the retrieval of incorrect papers that may have numerous citations, particularly if they are from popular fields of study—this could skew our analysis.

We have enhanced the NIME PA by incorporating a more robust algorithm that searches for papers in Semantic Scholar. This algorithm progressively queries the database with up to 12 different search strings, created from different combinations of the title, authors' last names, year of publication, and the string 'NIME'. The first result returned by the query is accepted as valid only if it matches the authors' names and number. If no matching paper is found, the NIME PA prompts the user to manually enter the paper's unique identifier from Semantic Scholar.

Currently, there are 26 NIME papers that require a manual search and specification of the correct paper identifier. The majority of these are from the 2023 and 2021 proceedings. Upon manually verifying all papers with more than 20 citations, we found 3 papers that we had to exclude from the search. This was due to an incorrect association with a significantly large number of citations in Semantic Scholar. Among these, one is a short paper and two are related to installations. We also inspected all papers that were not found and excluded 8 from the search; these were short plain text documents briefly describing an installation or a performance. This refined search approach has significantly reduced the number of papers not found in Semantic Scholar from 17.2% [6] to just 0.8%. Additionally, it has decreased the number of papers requiring a manual amendment of citation count from a couple of dozen to just one.

After identifying the correct paper, we use its unique identifier to perform a lookup in the Semantic Scholar database. This allows us to retrieve and incorporate the following data for each NIME paper into the tabular structure produced by the NIME PA:

- Unique identifier of the paper.
- Unique identifiers of all the authors.
- Too Long; Didn't Read (TLDR) short summaries [17].
- Scientific Paper Embeddings using Citation-informed Transformers (SPECTER) 768-dimensional vectors representing the papers [18].
- Number of references.
- Number of citations.
- Number of highly influential citations.
- List of references and list of citations with:
    - Title and unique identifier of each paper.
    - Names and unique identifiers of all its authors.
    - Publication year.
    - Publication venue and type.
    - Fields of study estimated by Semantic Scholar.

Semantic Scholar can only index papers for which a publicly accessible PDF is available. However, a paper without an attached PDF might still appear in the database. Yet, the information retrieved under these circumstances is likely to be incomplete or unreliable. In this study, particularly when analyzing references, we only consider data retrieved from NIME papers that include a non-empty list of references. The presence of a non-empty list of references suggests that the PDF is both available and properly analyzed.

The lists of references and citations are processed and consolidated into several more tabular data structures. Within these structures, each reference or citation is listed only once and includes all the information retrieved from Semantic Scholar, along with the following additions:

[2] https://github.com/NIME-conference/NIME-bibliography

[3] https://www.semanticscholar.org/

- Total count of times a paper has been cited or referenced.
- Year-over-year distribution of citing or referencing count.

These are further processed to create an additional tabular data structure, in which each author who is cited or referenced appears only once. The following details are included for each author:

- Name and unique identifier.
- Total count of citations and of references.
- Distribution of the citing or referencing count over the years.
- An indicator of whether the author has ever published in the NIME proceedings.

The citing and referencing counts for authors are processed individually; thus, if a paper has multiple authors, the tally increases for each one. These tabular data structures are then further intersected, merged, and mined to generate the figures and metrics presented within this paper. In most of our visualizations, we restrict the display to the top 40 entries from each reference or citation ranking. Comprehensive tabular data including full rankings and yearly breakdowns, along with the open-source code that produces the results presented in this paper, are available online[1]. All data presented in this paper are derived from the NIME proceedings archive and the Semantic Scholar database, mined on April 27, 2024.

## 2.1 Data Overview

As of the current date, the NIME proceedings corpus contains 2,110 papers published between 2001 and 2023. Within Semantic Scholar, we have computationally identified 2,069 of these papers, while an additional 26 were located with the help of manual searches. Together, these papers account for 99.2% of the entire corpus and form the basis of our analysis. However, 168 papers within this subset have been retrieved with incomplete reference-related information, likely due to the absence of publicly available PDFs. Consequently, for the purposes of reference analysis or body-text data extraction, such as obtaining the SPECTER embedding, our study is based on 1,926 papers, representing 91.3% of the total corpus. Figure 1 presents a yearly breakdown of these metrics, clearly indicating a discrepancy in the works published in 2021 and 2022; this is further explained in Section 5.1. For the remaining, the small number of papers not indexed in Semantic Scholar can be due to the inclusion of non-paper materials in the proceedings, such as installations, and performances.

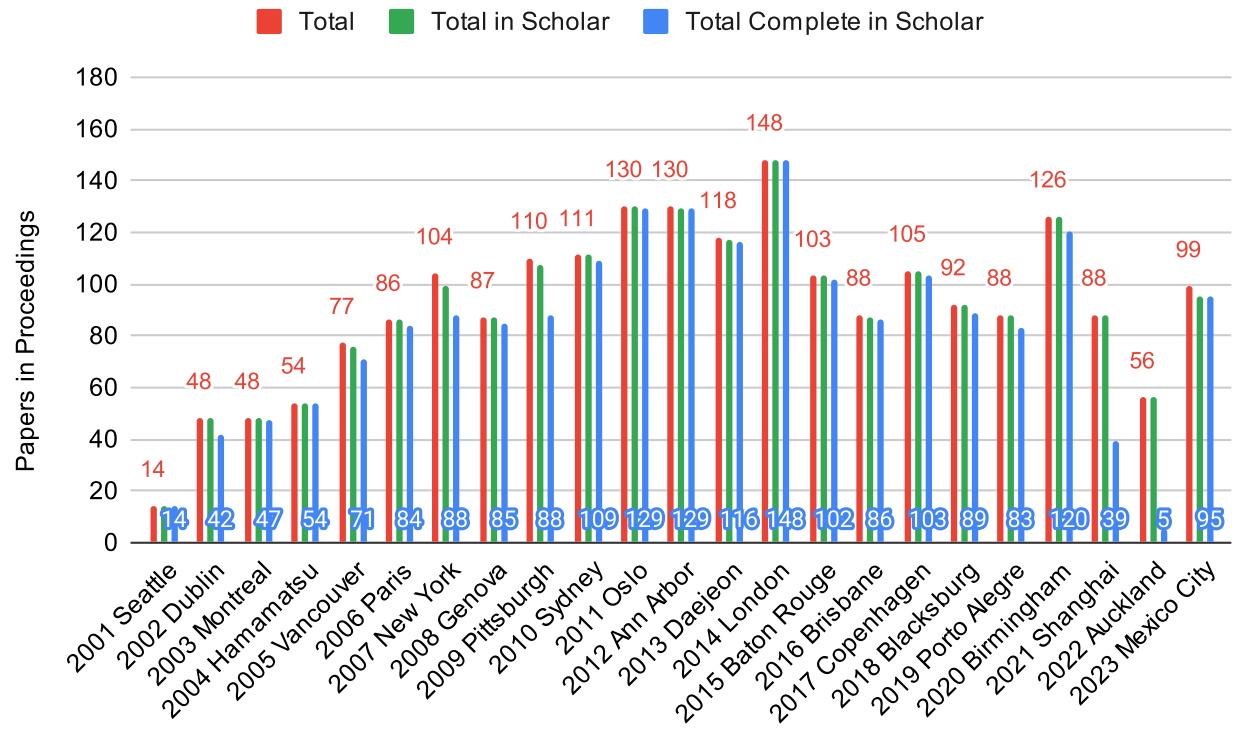

**Figure 1. Number of papers in the annual NIME proceedings (red), their indexing status in the Semantic Scholar (green), and papers indexed with complete reference data (blue).**

The 2,110 papers in the NIME proceedings feature a total of 5,325 authors, representing 2,841 unique individuals. Within Semantic Scholar, these papers are associated with 2,947 different author identifiers, suggesting that some individual authors have been assigned more than one profile. The corpus of papers contains 31,989 non-unique references, which averages to 16.6 references per paper. NIME papers have been cited 24,384 times, with an average of 11.6 citations per paper. However, as detailed subsequently and in our prior work, the citation distribution across the corpus is highly skewed [6]. The analysis of references and citations is discussed in greater detail in the following two sections.

# 3. REFERENCES ANALYSIS

A total of 12,935 unique works appear in the reference lists of NIME papers, totaling 31,989 references. These are distributed across the annual proceedings, as illustrated in Figure 2. The total number of non-unique references (orange) and the number of newly unique citations (blue, i.e., those not cited in previous NIME proceedings) correlate with the number of papers in the annual proceedings, as shown in Figure 1. However, the average number of references per paper (cyan) shows an upward trend in recent years, with a significant peak in 2023. The 2022 average, along with most reference-related figures from this year, should be disregarded as it is based solely on data from 5 papers.

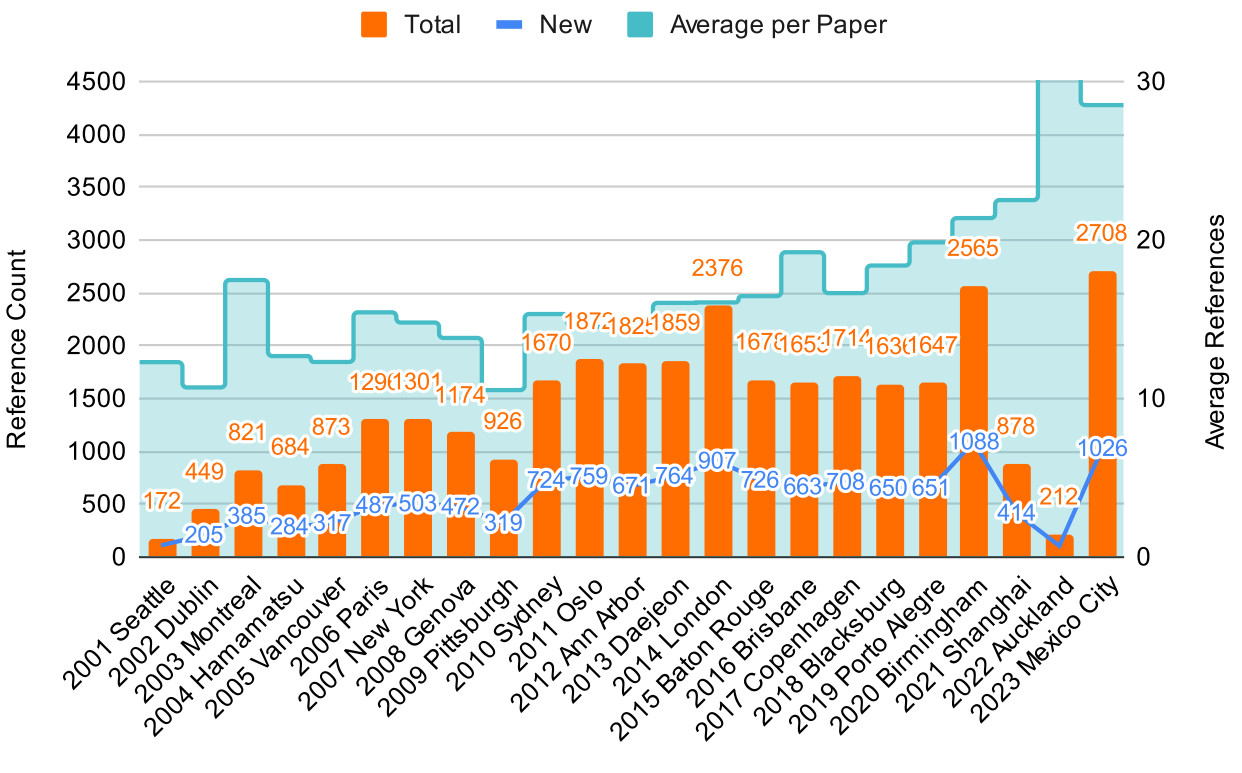

**Figure 2. Total (orange), new as not previously cited (blue) and average per paper (cyan) number of works appearing in the list of references in annual NIME proceedings.**

In assessing the extent to which the NIME publications are self-referential, we examined how often the works included in the list of references are part of the NIME proceedings corpus itself. On average, references to other NIME papers account for 17.3% of the total. As visible in Figure 3 (blue), this percentage has gradually increased over time in line with the growth of the proceeding's corpus. The nonzero percentage for 2001 can be attributed to some papers having been subsequently extended to journal versions and likely merged in the Semantic Scholar database. Comparatively, we also examine authors appearing in the lists of reference, and we determine whether they have ever authored or co-authored a NIME paper. This percentages rise to an average of 46.5% (red). These metrics do not strongly correlate with each other for the first ten annual proceedings, as shown in Figure 3. When calculating the latter, we did not consider whether the authors referenced had already published in NIME at the time of their first appearance in the list of references; that is, an author may have been cited before they published in the NIME proceedings.

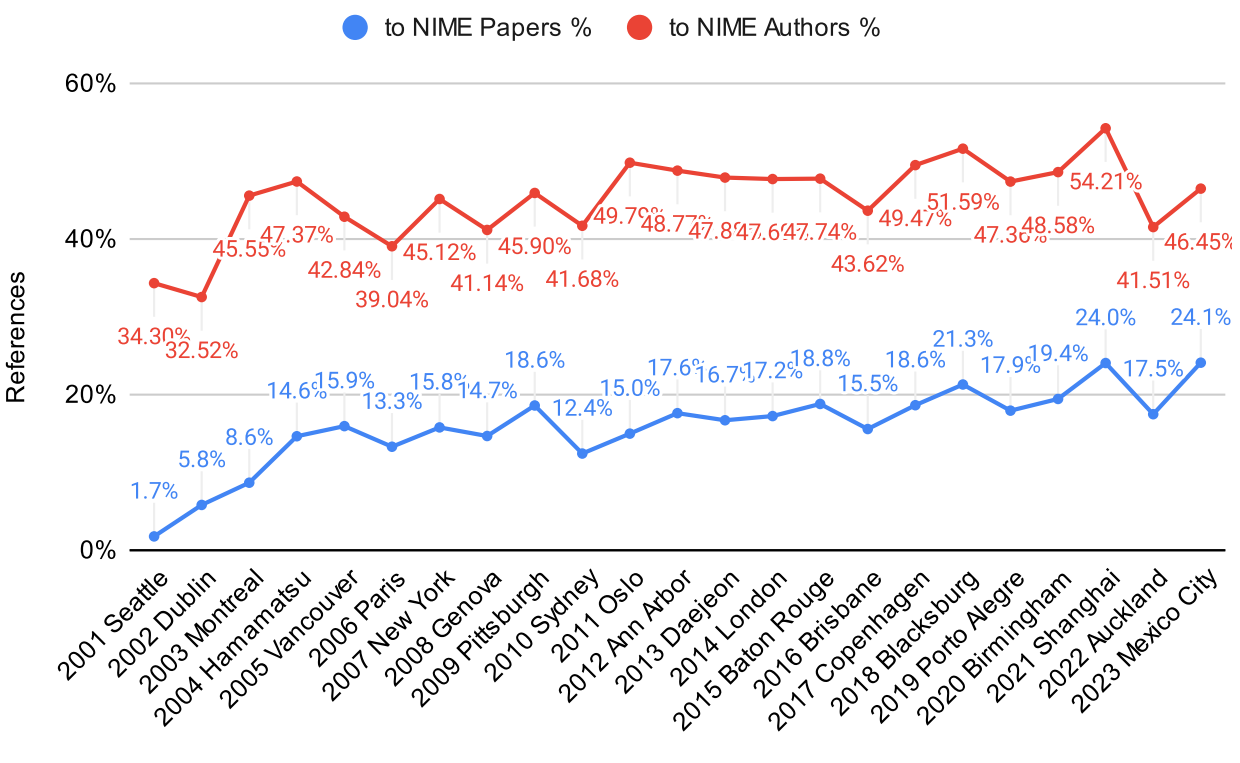

**Figure 3. Percentage of references in annual proceedings to other NIME papers (blue) and to works by authors that have authored at least one NIME paper (red).**

Figure 4 depicts a histogram showcasing the distribution of the 2,110 NIME papers plotted against the number of references. This display shows a Gaussian-like distribution with a mean of 16.6 and a standard deviation of 10.8.

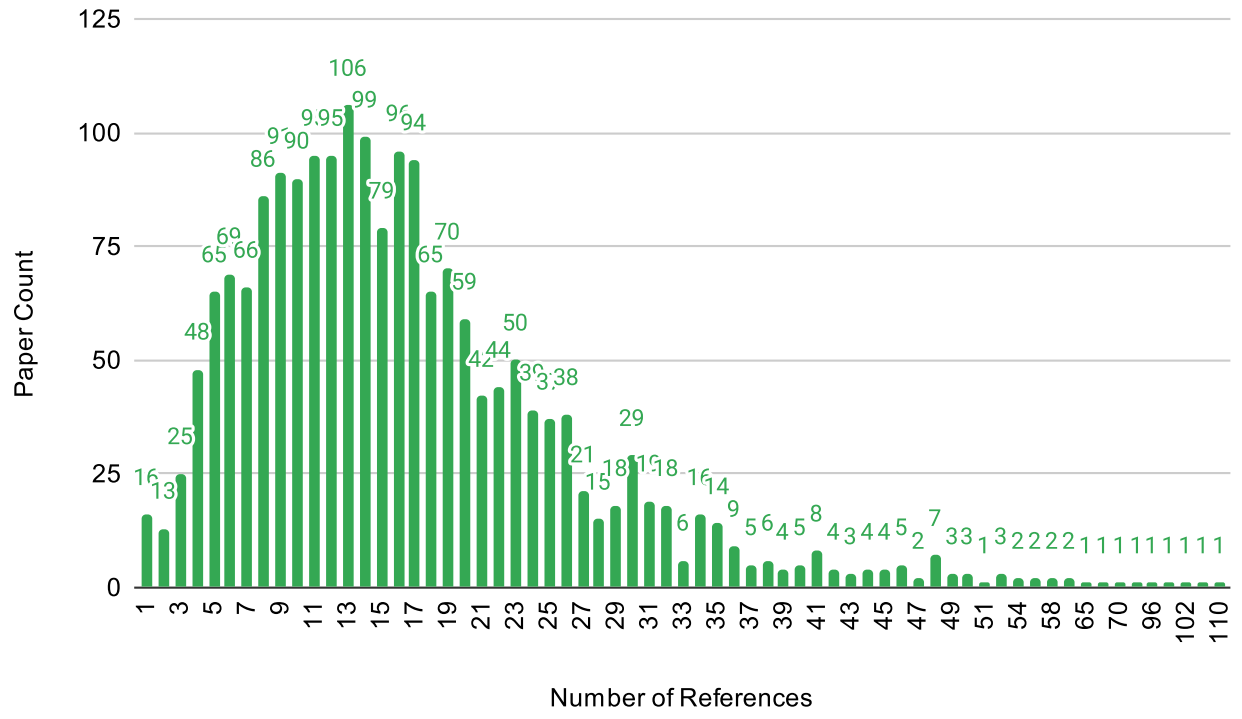

**Figure 4. Distribution NIME papers according to number works appearing in their lists of references.**

Figure 5 illustrates the distribution of the 31,989 non-unique references plotted against their relative ages, revealing an exponential-like distribution with an average age of 8.5 years. For this analysis, we defined the relative age as the difference between the publication year of the NIME paper and that of the referenced paper. As Figure 5 shows, 314 references appear to be more recent than the papers citing them. This discrepancy could be due to the referencing of preprints, extensions of works into journal versions, or inaccuracies within the Semantic Scholar database.

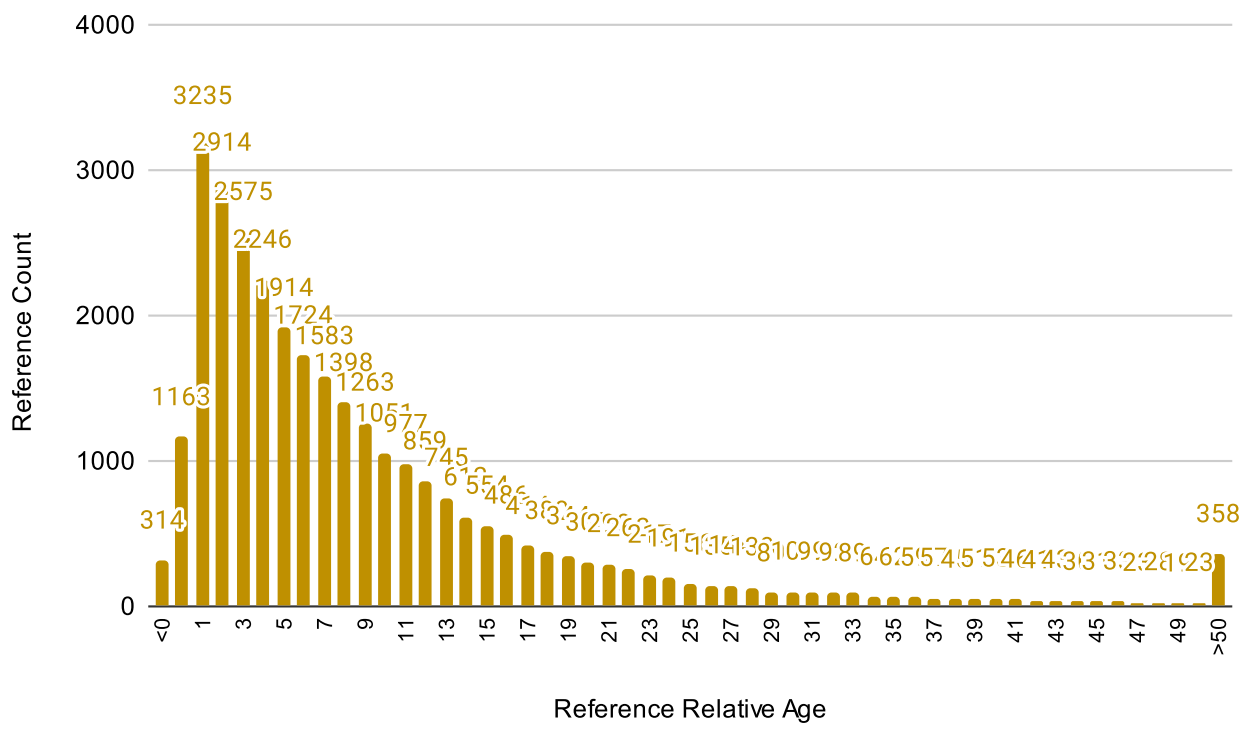

**Figure 5. Distribution of references appearing in NIME proceedings according to their relative age, defined as the difference between the publication year of the NIME paper and that of the referenced paper.**

Each paper in the Semantic Scholar database is automatically tagged with one or more broad fields of study, as determined by their machine learning models, which allows for 23 possible options. Figure 6 shows the distribution of tags across these fields. It displays the 31,989 non-unique references (cyan, left axis), the 5,519 non-unique references to other NIME papers (orange, right axis), and the 2,110 papers in the NIME proceedings (yellow, right axis). The tag 'Computer Science' is noticeably dominant, followed by 'Art' and 'Engineering'.

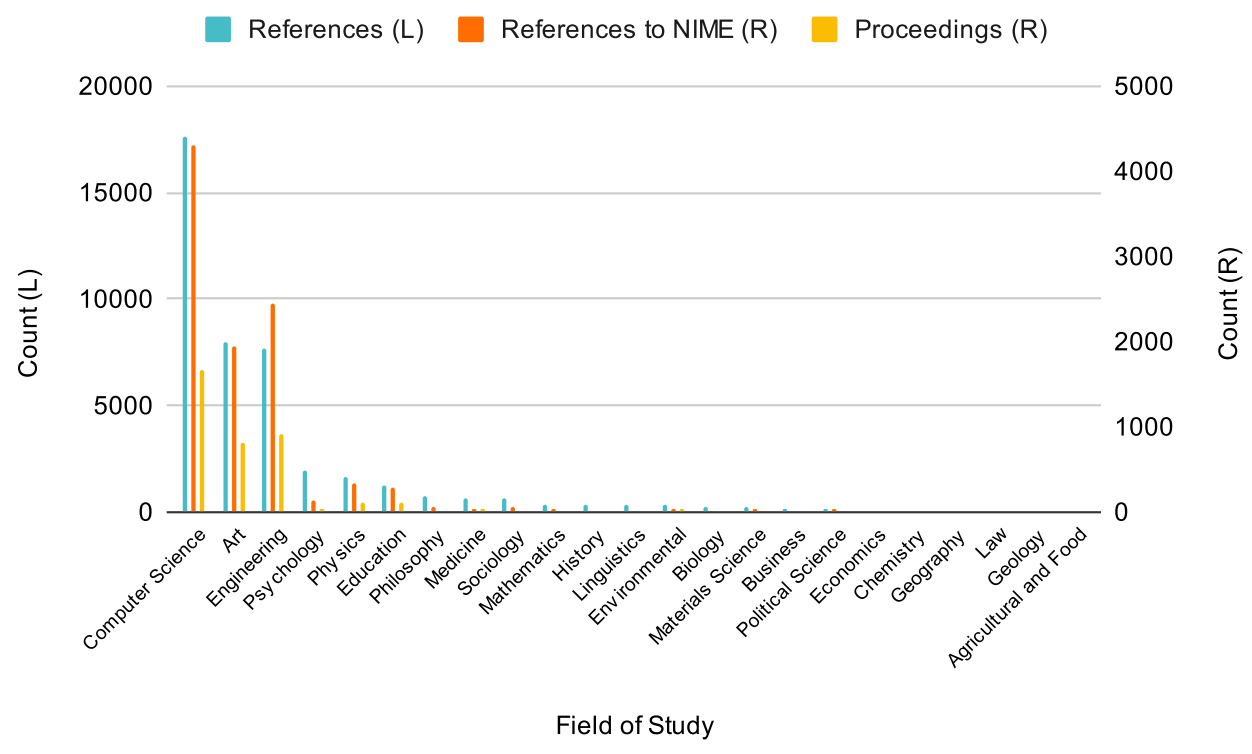

**Figure 6. Distribution of 'Field of Study' tags, as automatically assigned by Semantic Scholar, across three categories: works appearing in the reference lists of NIME papers (cyan, left axis), works appearing in the reference lists of NIME papers and belonging to the NIME proceedings corpus (orange, right axis), and works published in the NIME proceedings (yellow, right axis).**

The works appearing in the lists of references in NIME proceedings have been published in 1113 different scholarly venues, automatically detected by Semantic Scholar. Figure 7 details the distribution of non-unique references across the top 40 most popular venues, representing a total of 11,852 references, which account for 37.1% of those retrieved from Semantic Scholar with a valid 'publication venue' field. However, upon manual review of the data, we discerned that the publication venue often goes undetected, especially for relatively old papers, particularly those published in the Computer Music Journal (CMJ) and the International Computer Music Conference (ICMC). As a result, the actual counts for these venues are likely to be higher. Papers published in the book Trends in Gestural Control of Music were consolidated manually.

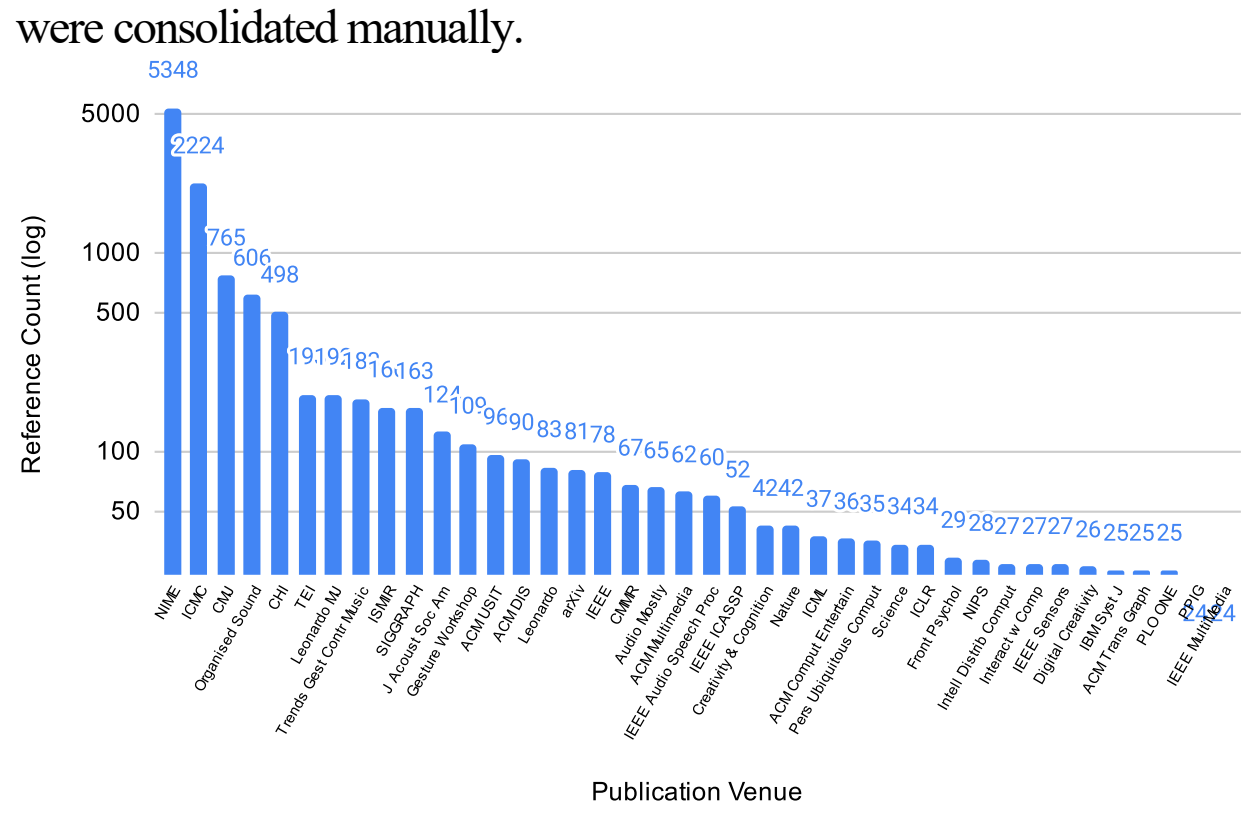

**Figure 7. Top 40 publication venues appearing most frequently in the lists of references of NIME proceedings.**

The works referenced in the NIME proceedings cite a total of 18,782 unique authors. Among these, 2,206 have also contributed to the NIME proceedings themselves, representing 74.8% of the total unique individuals who have been involved in authoring NIME papers. We tallied the number of references attributed to each individual, and Figure 8 showcases the top 40 authors who appear most frequently. An author's count increases regardless of whether they have authored a paper individually or co-authored with others. The cumulative count of references for the authors featured in Figure 8 amounts to 10,064, which represents 16.1% of the total 62,485 references attributed to all authors. Except for one, all the authors listed in Figure 8 have published at least one paper in the NIME proceedings. Among the top 200 referenced authors, there are only 22 who have not authored a work in the NIME proceedings.

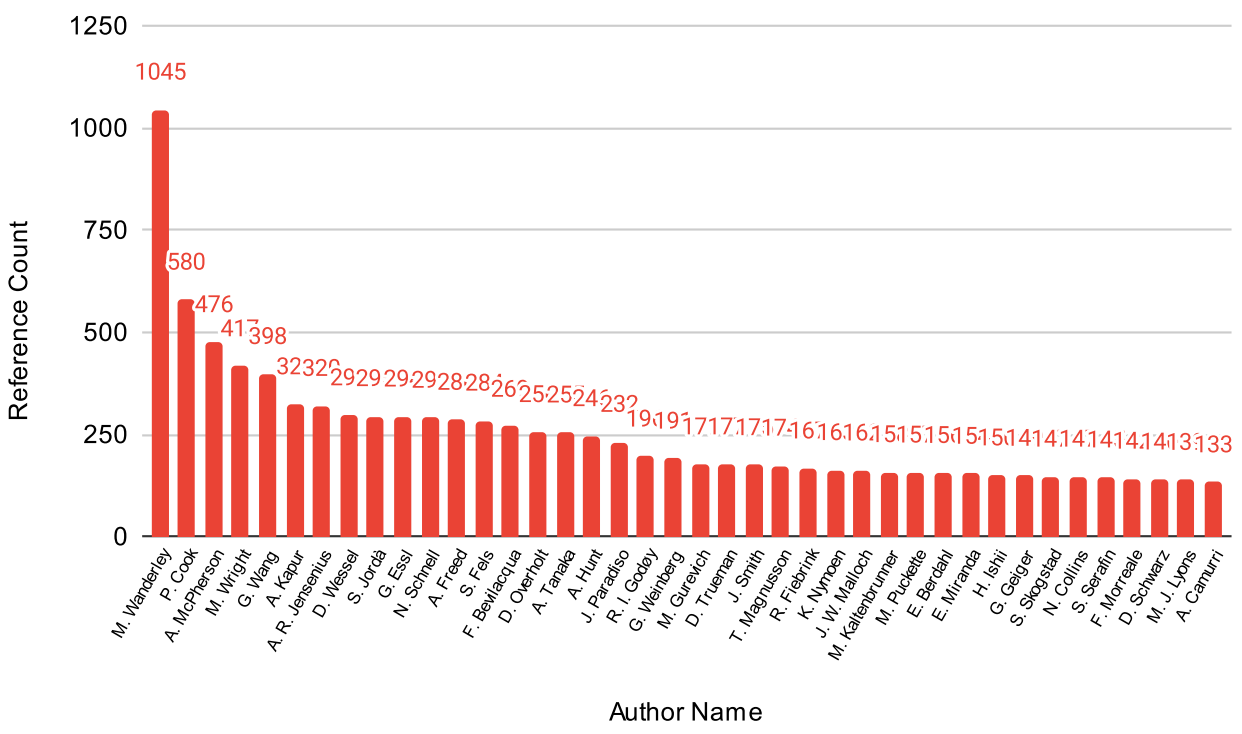

**Figure 8. Top 40 authors appearing most frequently in the lists of references of NIME proceedings.**

Table 1 lists the top 40 works most frequently appearing in the reference lists of NIME papers. The table provides the titles, publication venues, and years of publication for these works. The entries are sorted by the number of times they are cited in NIME papers, as shown in the 'Refs' column. To offer an additional metric that may reflect the overall influence of the work, we include the 'Refs-c' column, which tallies the total number of citations that referencing NIME papers have received. We have manually verified and corrected the publication venues where necessary. Together, these works comprise 2,367 of the 31,989 references found in NIME papers, which amounts to 7.4% of the total.

**Table 1. The top 40 works most frequently appearing in list of references of NIME proceedings, including number of appearances (Refs), total citations received by the referencing NIME papers (Refs-c), title, publication venue and year.**

| Refs | Refs-c | Title | Venue | Year |
|---|---|---|---|---|
| 122 | 2372 | Problems and Prospects for Intimate Musical Control of Computers | NIME/CMJ | 2002 |
| 98 | 1511 | Principles for Designing Computer Music Controllers | NIME | 2001 |
| 87 | 1672 | The Importance of Parameter Mapping in Electronic Instrument Design | NIME | 2002 |
| 76 | 1880 | Input Devices for Musical Expression: Borrowing Tools from HCI | NIME | 2001 |
| 74 | 1808 | Open SoundControl: A New Protocol for Communicating with Sound Synthesizers | ICMC | 1997 |
| 68 | 1305 | Pure Data: another integrated computer music environment | ICMC/Intercoll | 1996 |
| 57 | 522 | New Digital Musical Instruments: Control and Interaction Beyond the Keyboard | A-R Editions | 2006 |
| 50 | 421 | The reacTable: exploring the synergy between live music performance and tabletop tangible interfaces | TEI | 2007 |
| 49 | 704 | The 'E' in NIME: Musical Expression with New Computer Interfaces | NIME | 2006 |
| 49 | 704 | Mapping Strategies for Musical Performance | Trends Gest Contr Music | 2000 |
| 48 | 1047 | Mapping performer parameters to synthesis engines | Organised Sound | 2002 |
| 44 | 648 | Contexts of Collaborative Musical Experiences | NIME | 2003 |
| 39 | 1566 | Towards a Model for Instrumental Mapping in Expert Musical Interaction | ICMC | 2000 |
| 38 | 651 | Mapping transparency through metaphor: towards more expressive musical instruments | Organised Sound | 2002 |
| 37 | 667 | Audiopad: A Tag-based Interface for Musical Performance | NIME | 2002 |
| 37 | 379 | Gestural control of sound synthesis | IEEE Proceedings | 2004 |
| 35 | 1708 | Instrumental Gestural Mapping Strategies as Expressivity Determinants in Computer Music Performance | AIMI Intl Workshop | 1997 |
| 35 | 1234 | The Computer Music Tutorial | MIT Press | 1996 |
| 35 | 430 | OpenSound Control: State of the Art 2003 | NIME | 2003 |
| 34 | 568 | Rethinking the Computer Music Language: SuperCollider | CMJ | 2002 |
| 34 | 241 | Design for longevity: ongoing use of instruments from NIME 2010-14 | NIME | 2017 |
| 33 | 347 | An Environment for Submillisecond-Latency Audio and Sensor Processing on BeagleBone Black | AES Conv | 2015 |
| 32 | 787 | MnM: a Max/MSP mapping toolbox | NIME | 2005 |
| 32 | 323 | The reacTable: a tangible tabletop musical instrument and collaborative workbench | SIGGRAPH | 2007 |
| 31 | 824 | Mobile Music Making | NIME | 2004 |
| 30 | 1201 | BoSSA: The Deconstructed Violin Reconstructed | ICMC | 2000 |
| 30 | 420 | Displaced Soundscapes: A Survey of Network Systems for Music and Sonic Art Creation | Leonardo MJ | 2003 |
| 30 | 294 | Designing Constraints: Composing and Performing with Digital Musical Systems | CMJ | 2010 |
| 30 | 170 | A History of robotic Musical Instruments | ICMC | 2005 |
| 29 | 576 | Playing by feel: incorporating haptic feedback into computer-based musical instruments | PhD Thesis Stanford | 2001 |
| 29 | 416 | A Framework for the Evaluation of Digital Musical Instruments | CMJ | 2011 |
| 28 | 351 | Evolving The Mobile Phone Orchestra | NIME | 2010 |
| 28 | 238 | A Meta-Instrument for Interactive, On-the-Fly Machine Learning | NIME | 2009 |
| 27 | 228 | Of Epistemic Tools: musical instruments as cognitive extensions | Organised Sound | 2009 |
| 27 | 218 | LEMUR's Musical Robots | NIME | 2004 |
| 26 | 729 | The Synthesis ToolKit (STK) | ICMC | 1999 |
| 26 | 720 | Digital lutherie - crafting musical computers for new musics' performance and improvisation | PhD Thesis UPF | 2005 |
| 26 | 570 | Musical Performance Practice on Sensor-based Instruments | Trends Gest Contr Music | 2000 |
| 26 | 473 | The Hands: A Set of Remote MIDI-Controllers | ICMC | 1985 |
| 26 | 426 | The Laptop Orchestra as Classroom | CMJ | 2008 |

## 4. CITATIONS ANALYSIS

Papers published in the NIME proceedings have been cited 24,384 times in 10,534 unique works. Out of these citing works, 1,343 are NIME papers themselves, representing 12.7% of the total. The distribution of citations across the annual proceedings is illustrated in Figure 9, which also displays normalized citation counts. To normalize these counts, we performed two divisions: first, by the number of papers in that specific year's proceedings, and second, by the age of the proceedings. This normalization is an attempt to provide a fair representation of the impact of the corpus of NIME papers published each year. As depicted in Figure 9, the normalized citation count for the 2001 Seattle proceedings is particularly high. This collection includes only 14 papers (as shown in Figure 1) but they have been cited 1,541 times to date.

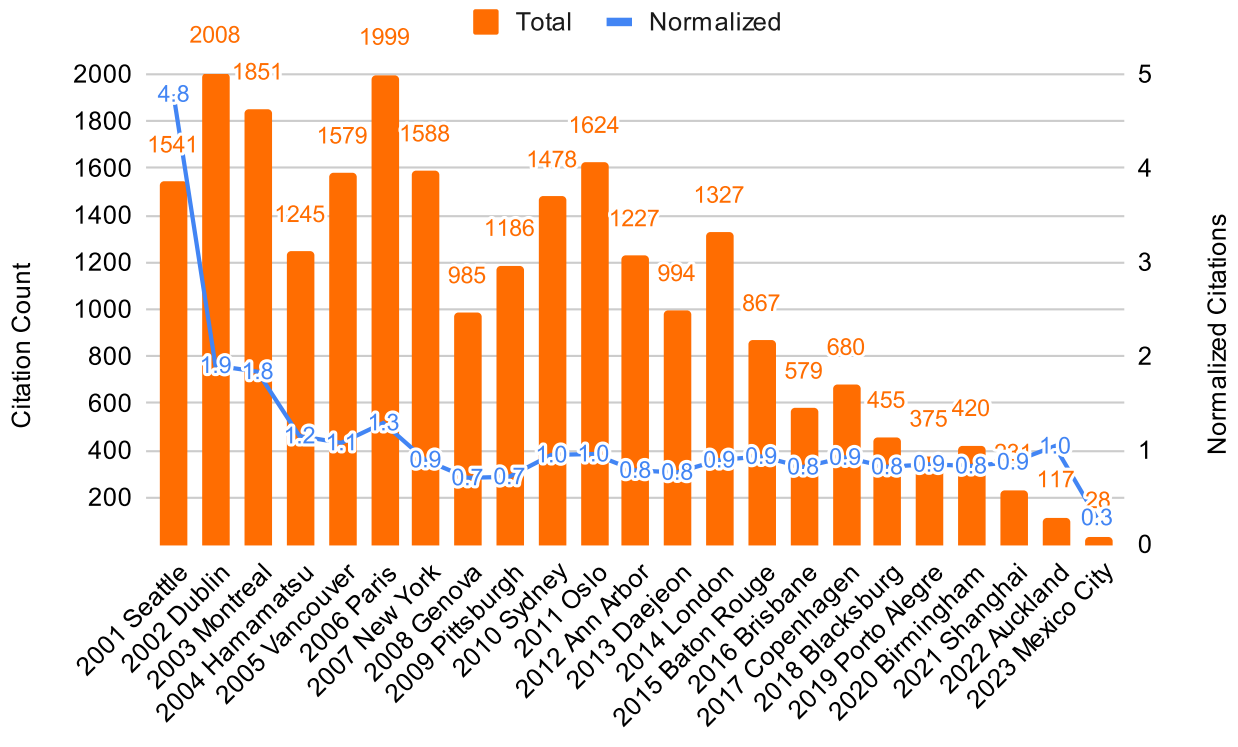

**Figure 9. Total (orange) and normalized (blue) number of citations to papers in the annual NIME proceedings. The normalization accounts for the size and age of the proceedings.**

Of the 24,384 citations received by NIME papers, 22.6% come from other works published in the NIME proceedings. However, when examining the authors who cite NIME papers and determining whether they have ever authored or co-authored a paper at NIME, the percentage jumps to 68.9%. The annual breakdown of these metrics is presented in Figure 10. In computing the latter percentage, as with the references analysis, we did not account for whether a citing author had published in NIME prior to making their citation; that is, an author may have published in NIME subsequent to citing a paper from the proceedings.

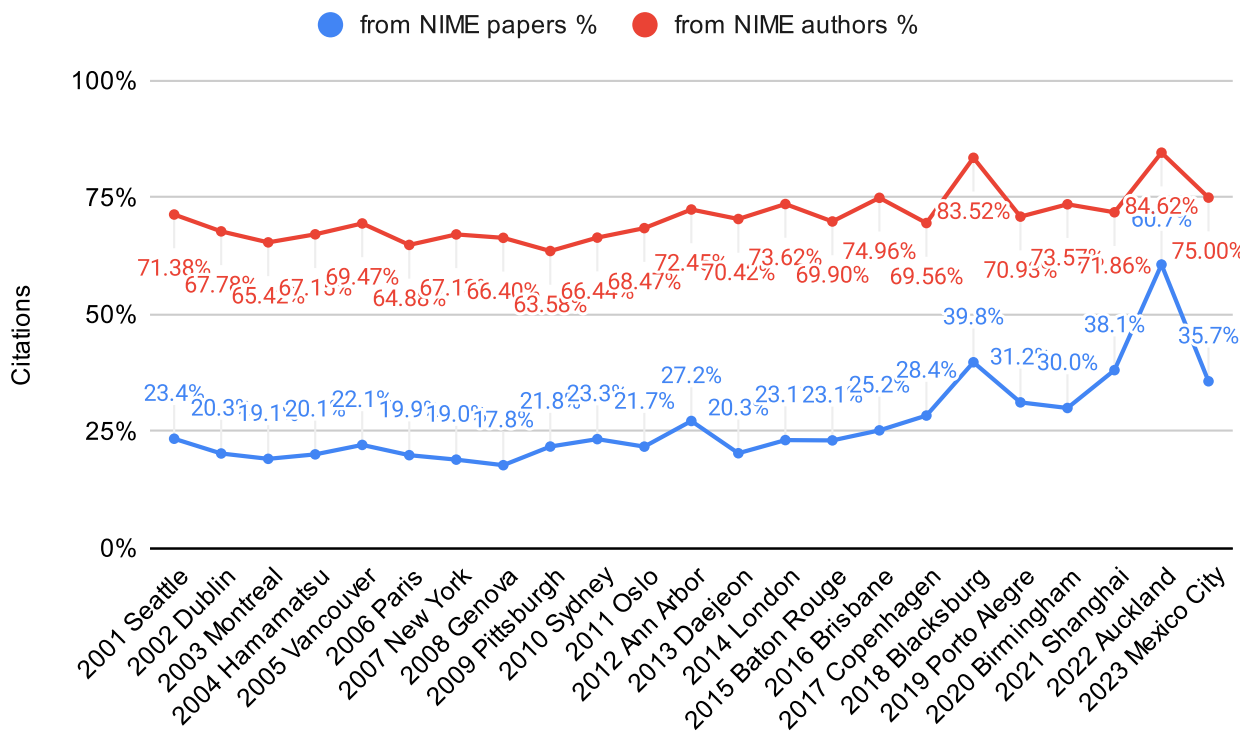

**Figure 10. Percentage of citations to papers in the annual proceedings from other NIME works (blue) and from authors that have authored at least one NIME work (red).**

The histogram in Figure 11 shows the distribution of the 2,110 NIME papers according to the number of citations they have received, revealing an exponential-like distribution with a mean of 11.6 and a standard deviation of 25.8. As is typical for most publication venues, the distribution is highly skewed. However, when evaluating other impact-related metrics [10], we find that only 331 NIME papers have never been cited so far (most of which are recent), accounting for 15.7% of the total. Notably, approximately half of the citations have been garnered by just 8.9% of NIME papers, indicating that a relatively small fraction of papers attracts a significant portion of the citations.

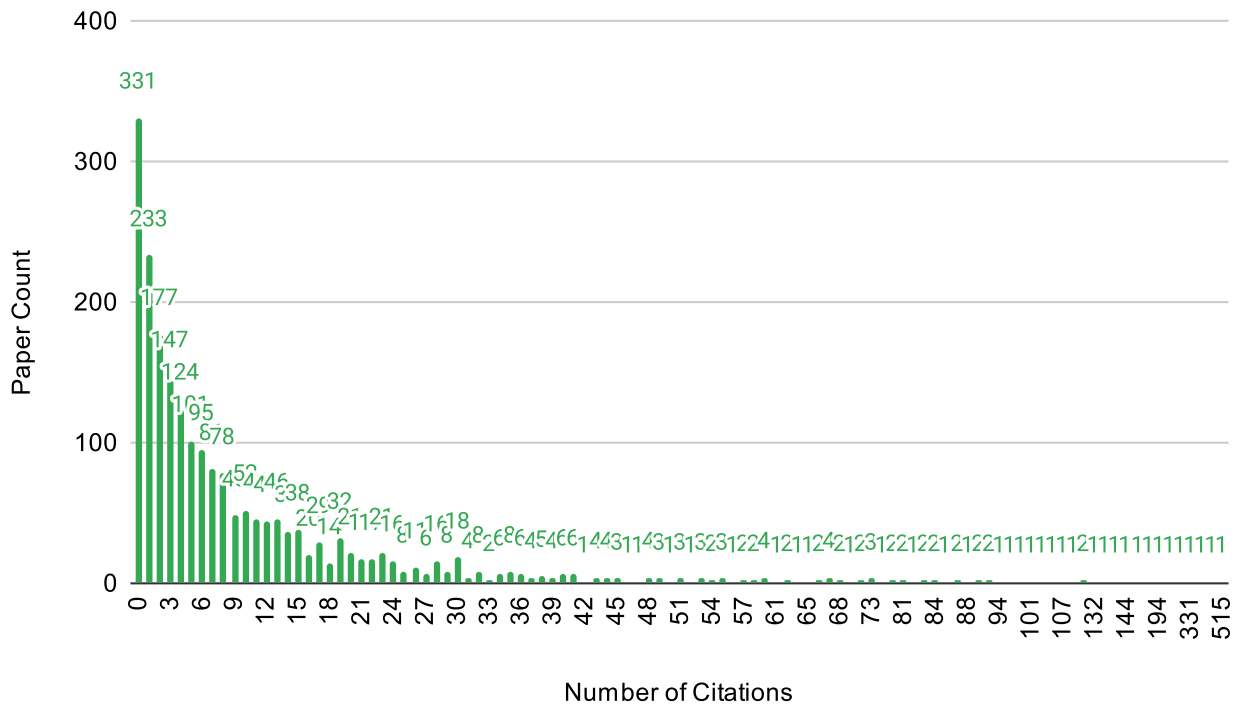

**Figure 11. Distribution NIME papers according to number of times they have been cited in other scholarly works.**

The distribution of the 24,384 non-unique citations against their relative age is shown in Figure 12, which presents a slightly exponential trend with an average of 6.1 years. Here, 'relative age' is defined as the difference in years between the publication of the citing paper and the cited NIME paper. As visible in Figure 12, there are 35 citations that appear more recent than the cited papers, likely due to inaccuracies in the Semantic Scholar indexing.

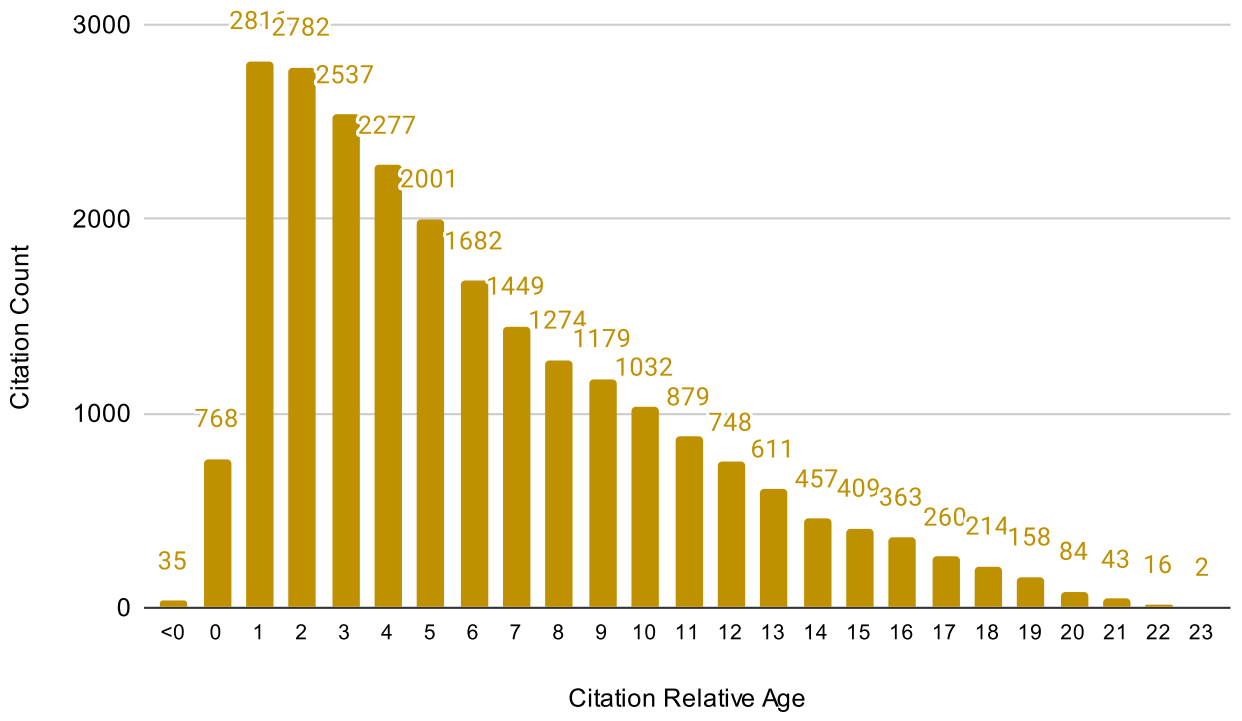

**Figure 12. Distribution of citations to works in NIME proceedings according to their relative age, defined as the difference between the publication year of the NIME paper and that of the citing paper.**

The distribution of the field of study tags for citations, as automatically inferred by Semantic Scholar, is depicted in Figure 13. This figure illustrates the distribution calculated for the 24,384 non-unique citing works (turquoise, left axis), for the 5,526 citations from other NIME papers (orange, right axis), and the 2,110 papers in the NIME proceedings (yellow, right axis). Also in this case the tag 'Computer Science' is noticeably dominant, followed by 'Art' and 'Engineering'.

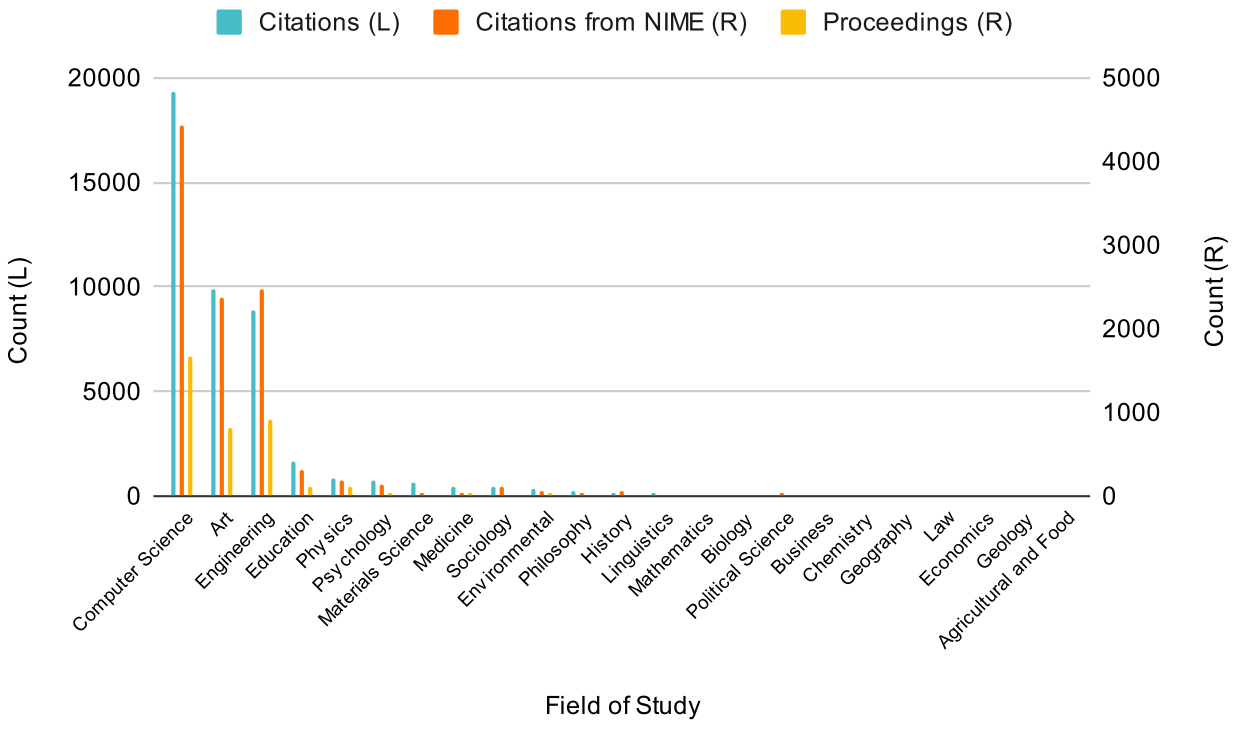

**Figure 13. Distribution of 'Field of Study' tags, as automatically assigned by Semantic Scholar, for works that cite NIME papers (cyan, left axis), works that cite NIME papers and belonging to the NIME proceedings corpus (orange, right axis), and works published in the NIME proceedings (yellow, right axis).**

Scholarly works that cite papers in NIME proceedings are from 976 different publication venues. The distribution across the 40 most popular venues is detailed in Figure 14, which accounts for a total of 10,156 citations. This represents 41.7% of the citations retrieved from Semantic Scholar that have a valid 'publication venue' field.

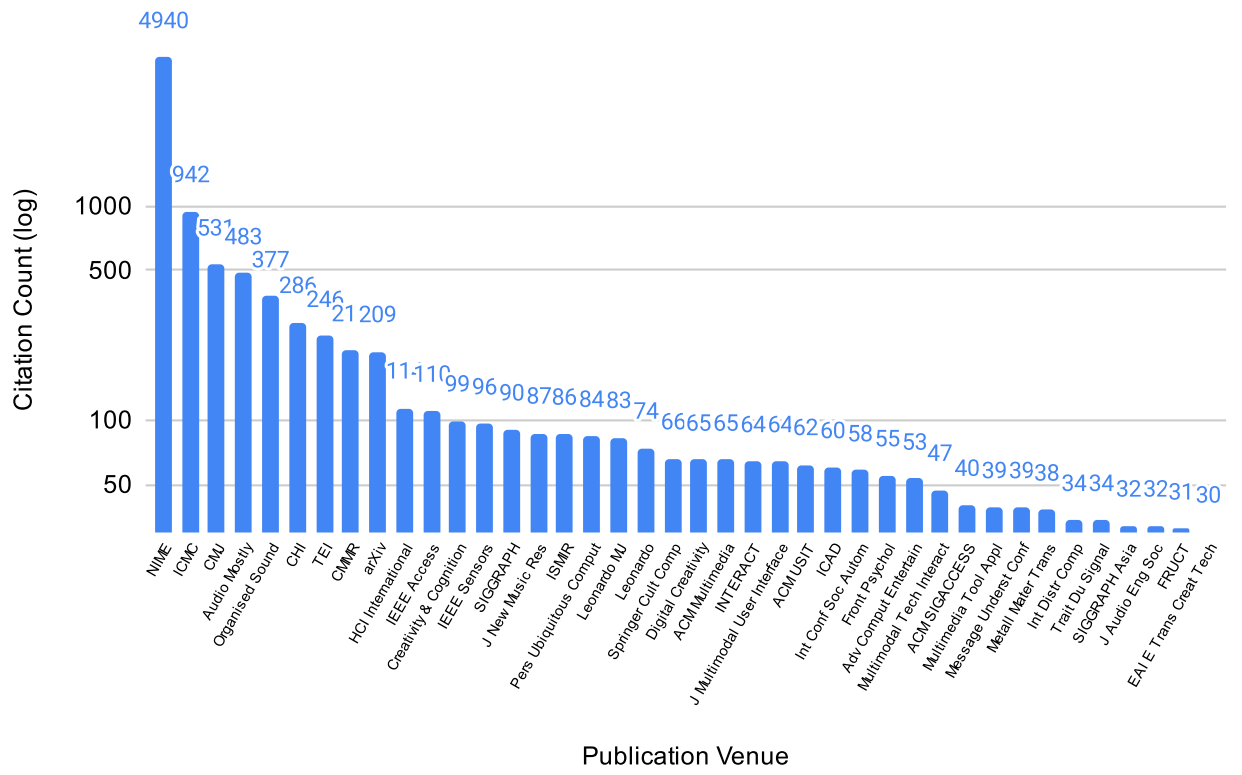

**Figure 14. Top 40 publication venues citing most frequently papers published in NIME proceedings.**

Works citing the NIME proceedings have been written by 16,032 unique authors, of which 2,394 are also contributors to the NIME proceedings, representing 81.2% of all NIME authors. We have tallied citations for each author, and the 40 most frequently citing authors are presented in Figure 15. The citation count reflects contributions both from solo-authored papers and from co-authored works. The combined total citations from authors shown in Figure 15 amount to 9,084, representing 13.5% of the 67,172 overall citations, which is the total if considering all citing authors. All authors featured in Figure 15 have published at least one paper in the NIME proceedings. Within the top 200 citing authors, only 13 have not authored a work in the NIME proceedings.

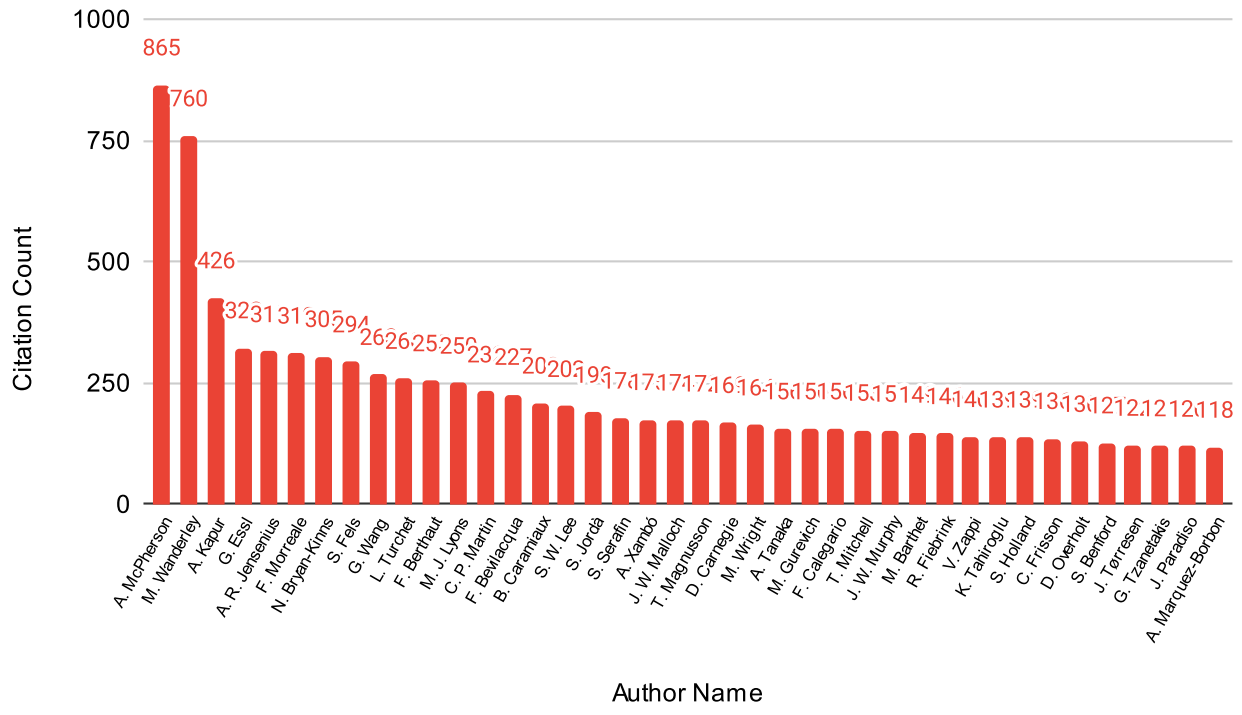

**Figure 15. Top 40 authors citing most works published in the NIME proceedings.**

When looking at works citing the largest collections of NIME paper, we find primarily doctoral theses in the top 40 positions. This is quite expected as theses are fairly comprehensive research work with extensive list of references compared to academic articles. For this reason, we have divided the top work citing most NIME papers in two tables, sorted by number of citations first and then by publication year. Table 2, includes the 40 top citing books, journal articles and conference papers, ranked according to the number of cited NIME and year (progressive). The table include number of cited papers, title, venue and year of publication. The publication venues have been manually checked and amended when needed. The works listed in Table 2 account for 1499 out of the 24,384 citations to NIME papers, equivalent to 6.1% of the total. PhD theses have been omitted from this table because their inclusion heavily skewed the results; these often cite a significant number of works, far more than a typical academic article. Additionally, we cannot ensure the comprehensive coverage of theses as not all universities maintain a public thesis archive with PDF files accessible by Semantic Scholar. A separate list of theses citing the largest collections of NIME papers is available online[1]. Course material and project reports have also been omitted, while book chapters have been consolidated under the volume in which they are included.

An in-depth analysis of the most-cited NIME papers and the distribution of citations among NIME authors is beyond the scope of this paper, as we already provided this in information in [6]. These figures can be easily updated with new data using the NIME PA.

**Table 2. The top 40 academic articles and books citing the largest collections of NIME papers with details on and number of cited papers (Cits), title, publication venue and year.**

| Cits | Title | Venue | Year |
|---|---|---|---|
| 71 | Machine Learning for Musical Expression: A Systematic Literature Review | NIME | 2023 |
| 66 | The O in NIME: Reflecting on the Importance of Reusing and Repurposing Old Musical Instruments | NIME | 2023 |
| 48 | New Directions in Music and Human-Computer Interaction | Springer | 2019 |
| 45 | Mobile Devices as Musical Instruments - State of the Art and Future Prospects | CMMR | 2017 |
| 42 | A Scale-Based Ontology of Digital Musical Instrument Desig | NIME | 2023 |
| 28 | Designing Digital Musical Instruments Using Probatio | Springer | 2019 |
| 27 | Surface Electromyography for Direct Vocal Control | NIME | 2020 |
| 26 | A Comprehensive Review of Sensors and Instrumentation Methods in Devices for Musical Expression | MDPI Sensors | 2014 |
| 26 | Discourse is critical: Towards a collaborative NIME history | NIME | 2021 |
| 22 | Design for longevity: ongoing use of instruments from nime 2010-14 | NIME | 2017 |
| 22 | Smart Musical Instruments: Vision, Design Principles, and Future Directions | IEEE Access | 2019 |
| 22 | Algorithmic Pattern | NIME | 2020 |
| 22 | Leveraging Android Phones to Democratize Low-level Audio Programming | NIME | 2023 |
| 20 | Interaction musicale numérique | Traitement du Signal | 2015 |
| 20 | Simple mappings, expressive movement: a qualitative investigation into the end-user mapping design of experienced mid-air musicians | Digital Creativity | 2018 |
| 20 | Deep Predictive Models in Interactive Music | arXiv | 2018 |
| 20 | Studying How Digital Luthiers Choose Their Tools | CHI | 2022 |
| 19 | Some reflections on the relation between augmented and smart musical instruments | Audio Mostly | 2018 |
| 19 | NIME Identity from the Performer's Perspective | NIME | 2018 |
| 19 | A NIME Of The Times: Developing an Outward-Looking Political Agenda For This Community | NIME | 2020 |
| 19 | Transmitting Digital Lutherie Knowledge: The Rashomon Effect for DMI Designers | NIME | 2023 |
| 18 | Vocal Control of Sound Synthesis Personalized by Unsupervised Machine Listening and Learning | CMJ | 2018 |
| 18 | Musicking with an interactive musical system: The effects of task motivation and user interface mode on non-musicians' creative engagement | Int J Human-Comp Stud | 2019 |
| 18 | Non-Rigid Musical Interfaces: Exploring Practices, Takes, and Future Perspective | NIME | 2020 |
| 18 | Towards a Latin American NIME Research Community | SBC | 2021 |
| 18 | PhonHarp: A Hybrid Digital-Physical Musical Instrument for Mobile Phones Exploiting the Vocal Tract | Audio Mostly | 2021 |
| 17 | Contexts of Collaborative Musical Experiences | NIME | 2003 |
| 17 | Music in Extended Realities | IEEE Access | 2021 |
| 17 | Sonic Interactions in Virtual Environments | Springer | 2022 |
| 17 | Raw Data, Rough Mix: Towards an Integrated Practice of Making, Performance and Pedagogy | NIME | 2023 |
| 16 | Designing Digital Musical Interactions in Experimental Contexts | NIME | 2011 |
| 16 | A Manual Actions Expressive System (MAES) | Organised Sound | 2013 |
| 16 | Supporting Non-Musicians? Creative Engagement with Musical Interfaces | Creativity & Cognition | 2017 |
| 16 | Culture and Politics of Machine Learning in NIME: A Preliminary Qualitative Inquiry | NIME | 2023 |
| 16 | Redesigning the Chowndolo: a Reflection-on-action Analysis to Identify Sustainable Strategies for NIMEs Design | NIME | 2023 |
| 15 | A Framework for the Evaluation of Digital Musical Instruments | CMJ | 2011 |
| 15 | Live Coding The Mobile Music Instrument | NIME | 2013 |
| 15 | Data-Driven Analysis of Tiny Touchscreen Performance with MicroJam | CMJ | 2019 |
| 15 | Electronic_Khipu_: Thinking in Experimental Sound from an Ancestral Andean Interface | CMJ | 2020 |
| 15 | Hands-Free Accessible Digital Musical Instruments: Conceptual Framework, Challenges, and Perspectives | IEEE Access | 2020 |

Figure 16 displays a scatter plot of the SPECTER embeddings of NIME papers, which have been reduced to two dimensions using t-distributed Stochastic Neighbor Embedding (t-SNE) with a perplexity of 25. Each solid circle in the plot represents a NIME paper. The color indicates the publication year and the size is proportional to the number of times the paper has been cited. Even though there's no specific meaning attached to the two components of this abstract representation (it is an embedding of an embedding), they allow for the identification of potential clusters or patterns. While t-SNE is a heuristic method, the results are consistent when the process is repeated. Notably, the most cited NIME works appear to cluster towards the center-right of the plot, similarities in their contributions. Works published between 2009 and 2013 are spread evenly across the space, whereas more recent works densely populate the left side of the plot. This distribution may offer insights that for further investigation into the trends of topics within the annual proceedings. Data from 2021 to 2023 have been excluded, as SPECTER data retrieved from Semantic Scholar for these years appeared to be based primarily on the title or abstract due to missing indexing or unavailable PDFs. This generated a very tight cluster that interfered with the rest of the data during the t-SNE computation.

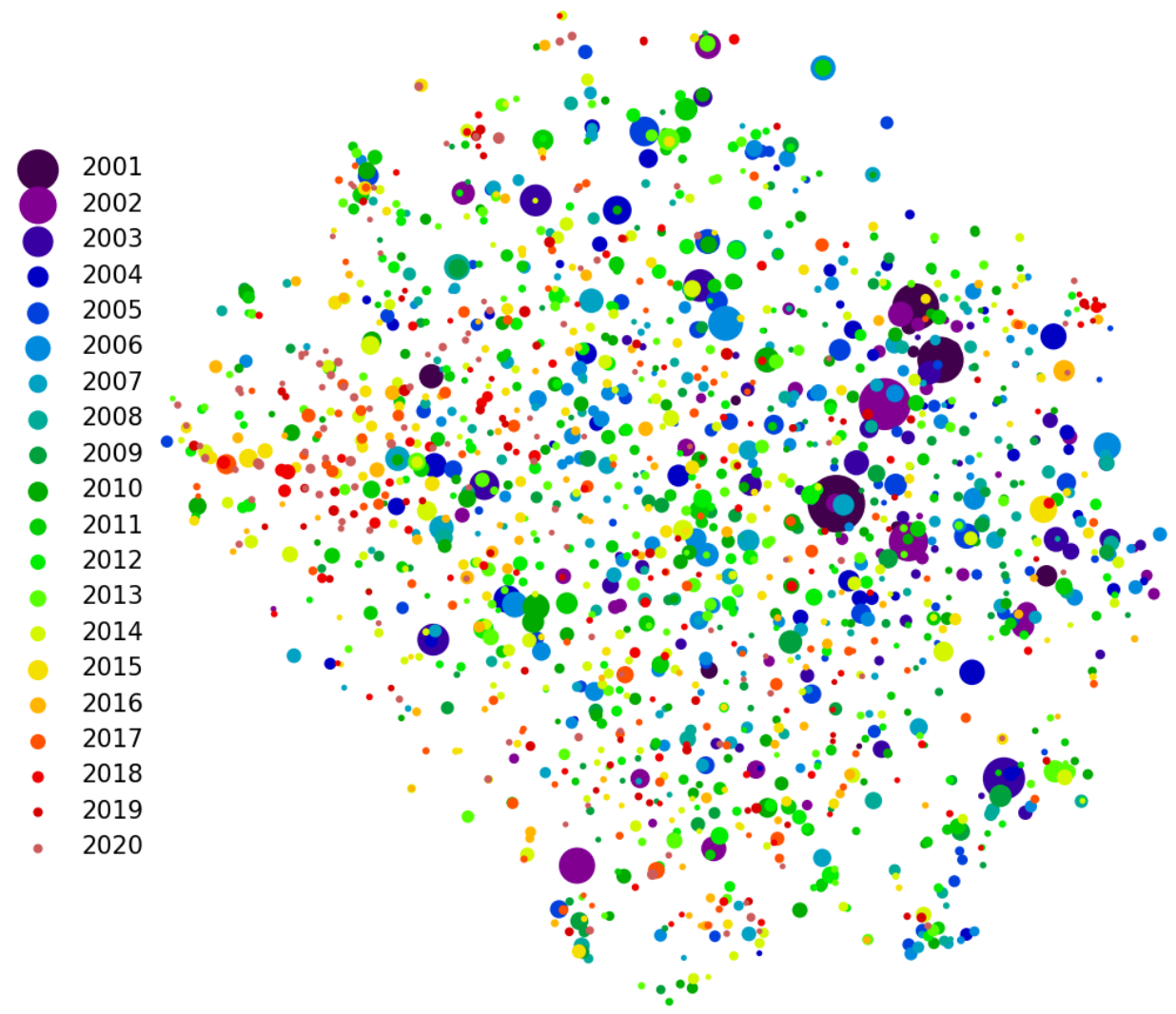

**Figure 16. Scatter plot of the SPECTER embedding of NIME papers up to 2020, reduced to two dimensions using t-SNE. Each solid circle represents a NIME paper; the color indicates the publication year, and the size is proportional to the number of citations received.**

## 5. DISCUSSION

The figures presented in the previous sections are not comparable to results from other publication venues within the field of technologies for sonic art and music, as no similar study has been conducted. Nonetheless, our open-source tool allows for the replication of this analysis for other venues, requiring only the BibTeX file containing all the works from conference proceedings or journal issues.

The percentages of references from and citations to other NIME papers, detailed in Figures 3 and 10, suggest that, as expected, a significant body of work on musical interfaces is published in other venues. In examining the extent to which NIME is a self-citing publication venue, the rate is relatively low and below 20%, though it is gradually increasing as the NIME corpus of papers expands. This rate is consistent with those of major journals [19]. However, almost 70% of the works that cite NIME papers are authored by individuals with some connection to NIME, indicating that the NIME community tends to look inward [4]. Nonetheless, there are positive aspects to self-citations within a publication venue, particularly when citing recent works, a trend that is common in high-impact journals and does not affect the resulting impact factor [19]. Indeed, Figure 5 reveals a tendency in NIME papers to use as references predominantly recent publications. This tendency may also reflect authors staying current with the latest techniques and technologies; however, it might also suggest a tendency towards superficial literature reviews that overlook seminal research on musical interfaces initiated in the mid-1970s [5]. Additionally, Figure 12 demonstrates that NIME papers are more likely to be cited within the first five years after publication. Beyond this period, papers may become less relevant, possibly because they primarily highlight design practices that quickly become dated, rather than emphasizing enduring design sciences.

'Computer Science', 'Art', and 'Engineering' are the three leading fields of study represented in NIME papers, as well as in their references and citations. This is evident in Figures 6 and 13. Approximately 55.1% of references and 79.5% of citations carry the 'Computer Science' tag. In contrast, 'Art' is included in 24.8% of references and 40.6% of citations, while 'Engineering' figures in 24.1% of references and 36.3% of citations. These patterns, also present in the NIME proceedings' tags, indicate that the field's multidisciplinary nature tilts heavily toward the applied sciences. Other, less frequent tags such as 'Psychology', 'Physics', and 'Education' are present in references and citations but less so in the proceedings themselves. It's important to note that these tags can overlap, as one paper may have multiple tags. The cumulative percentage of tag occurrences is 139.9% for references and 181.1% for citations, indicating a broader multidisciplinary scope in the latter.

Upon further investigating these trends across the annual proceedings, we observed that while the diversity (normalized standard deviation) of reference tags has remained fairly constant, the diversity of citation tags has been gradually decreasing since 2013. In contrast, the diversity of tags in NIME proceedings increased steadily until 2014, after which there was a slight decline and a plateau. These findings seem to contradict recent efforts by the NIME community to broaden its perspective and enhance diversity. However, it's possible that the effects of newly implemented initiatives have not yet become evident. Our analysis of references to, and citations from, other NIME works indicates similar trends, with 'Engineering' representing a larger share in both references and citations. This suggests that within the NIME community, works are particularly influential for research with a technical emphasis.

As expected, a significant proportion of NIME references have been published in the proceedings of the International Computer Music Conference (ICMC) and in the Computer Music Journal (CMJ), where most musical interface works were published in the pre-NIME era [5]. Other related publication venues include Organised Sound, Leonardo Music Journal, the Conference of the International Society for Music Information Retrieval (ISMIR), the International Symposium on Computer Music and Multidisciplinary Research (CMMR), and Audio Mostly. Older publication venues are more likely to feature among the most referenced in Figure 7. However, we also find Trends in Gestural Control of Music in a top position. Despite being a single volume published in 2000, it includes numerous scholarly works on musical interfaces, many of which have been highly influential in the early NIME community [5]. Regarding other types of publication venues, most relate to computer science and engineering subfields such as human-computer interaction, user interfaces, interaction techniques, signal processing, multimedia, acoustics, and technologies for creative arts.

In Figure 14, NIME citation publication venues appear to be more evenly spread than references, with ICMC and CMJ collecting a considerably lower percentage of all citations. This could be due to the emergence of specialized conferences in the field of sonic art and music technologies over the last two decades. Additional related publication venues listed include the Journal of New Music Research, the International Community for Auditory Display (ICAD), and the Journal of the Audio Engineering Society (AES). Among the top 40 publication venues that most frequently cite NIME papers, as shown in Figure 14, 22.9% of citations come from publications not directly related to sound and music technologies. This, in our opinion, is a positive metric, demonstrating the relevance of NIME research in non-creative or non-musical contexts. The good quality of NIME research is also suggested by the relatively low percentage of uncited papers, which stands at 15.6%, including recent works. This percentage is considerably lower than those in other disciplines except the medical field [15]. In general, figures suggests a constant increase in research on musical interfaces, not exclusively presented at NIME, and in the number of scholars in the field.

While assessing frequently referenced and citing authors, it's important to note that the data may not offer a complete picture. The method of counting only first authors, as we accomplished in [6], may result in different figures, and neither approach may fully capture individual contributions. The impact of contributions can significantly vary even when adhering to Vancouver authorship recommendations. Some publication platforms, primarily journals, have begun to offer standardized options to specify each author's contribution. The wider introduction of such practices could greatly benefit research, and the NIME community should consider evaluating or experimenting with this approach for future proceedings.

Table 1 presents the top 40 most referenced works, which include well-known works on theories, studies, reviews, and tools related to NIME, and well known musical interfaces. Most of these works were published before or during the first years of the NIME conference. Conversely, the works that cite the most NIME papers are primarily composed of a range of recent NIME-related studies, as shown in Table 2. Here we provide only a summary of the metrics and figures obtained through computational analysis—more in-depth assessments focusing on specific trends related to papers, fields, venues, authors, or their groups across annual proceedings can be conducted easily using the data we've compiled and shared online[1]. Also, NIME researchers can further exploit the data we shared to uncover lesser-known NIME studies that may be less visible as published in venues they do not usually consult, or using different keywords and terminology, hence not showing in academic search engines. Additionally, NIME researchers can use the data we provided to discover lesser-known NIME-related works that may be less visible because they are published in collections they generally do not read, or utilizing different keywords and terminology that do not readily appear in academic search engine results. As an example of such mining, works appearing in both the NIME references and citations are likely relevant to the NIME field. We found 2,561 of such works and 991 of them are not part of the NIME proceedings. Examining the latter may reveal previously unknown literature. Table 3 lists 40 of those works, ranked by the combined count of their appearances in the reference lists of NIME papers and the number of times they cite NIME papers. In this case as well, PhD theses have been excluded.

**Table 3. The top 40 non-NIME articles and books that are both referenced in and cited by NIME papers, detailing appearances in the reference lists of NIME proceedings (Refs), the number of times they cite NIME papers (Cits), title, publication venue and year; entries are ranked by the sum of Refs and Cits.**

| Refs | Cits | Title | Venue | Year |
|---|---|---|---|---|
| 50 | 2 | The reacTable: exploring the synergy between live music performance and tabletop tangible interfaces | TEI | 2007 |
| 48 | 1 | Mapping performer parameters to synthesis engines | Organised Sound | 2002 |
| 8 | 48 | New Directions in Music and Human-Computer Interaction | Springer | 2019 |
| 3 | 45 | Mobile Devices as Musical Instruments - State of the Art and Future Prospects | CMMR | 2017 |
| 37 | 7 | Gestural control of sound synthesis | IEEE Proceedings | 2004 |
| 29 | 15 | A Framework for the Evaluation of Digital Musical Instruments | CMJ | 2011 |
| 38 | 5 | Mapping transparency through metaphor: towards more expressive musical instruments | Organised Sound | 2002 |
| 26 | 14 | Open Sound Control: an enabling technology for musical networking | Organised Sound | 2005 |
| 33 | 6 | An Environment for Submillisecond-Latency Audio and Sensor Processing on BeagleBone Black | AES Conv | 2015 |
| 32 | 5 | The reacTable: a tangible tabletop musical instrument and collaborative workbench | SIGGRAPH | 2006 |
| 30 | 6 | Designing Constraints: Composing and Performing with Digital Musical Systems | CMJ | 2010 |
| 10 | 26 | A Comprehensive Review of Sensors and Instrumentation Methods in Devices for Musical Expression | MDPI Sensors | 2014 |
| 30 | 5 | A History of robotic Musical Instruments | ICMC | 2005 |
| 30 | 1 | Displaced Soundscapes: A Survey of Network Systems for Music and Sonic Art Creation | Leonardo MJ | 2003 |
| 2 | 28 | Designing Digital Musical Instruments Using Probatio | Springer | 2019 |
| 26 | 3 | The Laptop Orchestra as Classroom | CMJ | 2008 |
| 27 | 1 | Of Epistemic Tools: musical instruments as cognitive extensions | Organised Sound | 2009 |
| 20 | 8 | Interactivity for Mobile Music-Making | Organised Sound | 2009 |
| 14 | 13 | Collaborative Musical Experiences for Novices | J New Music Res | 2003 |
| 24 | 3 | Interconnected Musical Networks: Toward a Theoretical Framework | CMJ | 2005 |
| 19 | 8 | Do Mobile phones Dream of Electric Orchestras? | ICMC | 2009 |
| 17 | 8 | Digital Musical Interactions: Performer–system relationships and their perception by spectators | Organised Sound | 2011 |
| 18 | 6 | Audience-Participation Techniques Based on Social Mobile Computing | ICMC | 2011 |
| 2 | 22 | Smart Musical Instruments: Vision, Design Principles, and Future Directions | IEEE Access | 2019 |
| 9 | 14 | Instruments and Players: Some Thoughts on Digital Lutherie | J New Music Res | 2004 |
| 11 | 12 | An enactive approach to the design of new tangible musical instruments | Organised Sound | 2006 |
| 20 | 3 | The reacTable*: A Collaborative Musical Instrument | IEEE WETICE | 2006 |
| 22 | 1 | Why a laptop orchestra? | Organised Sound | 2007 |
| 21 | 2 | Continuous Realtime Gesture Following and Recognition | Gesture Workshop | 2009 |
| 22 | 1 | Stanford Laptop Orchestra (SLOrk) | ICMC | 2009 |
| 3 | 20 | Deep Predictive Models in Interactive Music | arXiv | 2018 |
| 10 | 13 | Accessible Digital Musical Instruments - A Review of Musical Interfaces in Inclusive Music Practice | MDPI Mmodal Tech Inter | 2019 |
| 21 | 1 | Composing for Laptop Orchestra | CMJ | 2008 |
| 16 | 6 | The Problem of the Second Performer: Building a Community Around an Augmented Piano | CMJ | 2012 |
| 18 | 3 | Gestural Control of Music | Workshop Eng & Music | 2001 |
| 16 | 5 | Toward Robotic Musicianship | CMJ | 2006 |
| 16 | 5 | The Machine Orchestra: An Ensemble of Human Laptop Performers and Robotic Musical Instruments | CMJ | 2011 |
| 18 | 3 | Advancements in Actuated Musical Instruments | Organised Sound | 2011 |
| 13 | 8 | Ocarina: Designing the iPhone's Magic Flute | CMJ | 2014 |
| 9 | 12 | Virtual Reality Musical Instruments: State of the Art, Design Principles, and Future Directions | CMJ | 2016 |

## 5.1 Issues in the NIME proceedings archival

The data presented in this paper was computationally extracted from the NIME proceedings archive. Consequently, any errors or inconsistencies in the archive could potentially propagate down to the results. Some of these inconsistencies were known in advance, while others were discovered during the ongoing process of developing our software tools for mining the NIME proceedings archive [16], which we used to compute the figures presented in this and our previous work [6]. During this process, we reported minor issues, such as malformed PDFs, typos in the records, or missing files, to the archive maintainers, who have since addressed them. Unresolved issues were handled through specific branches in the code. However, the issues discussed next persist and we recommend the NIME committee and community to reflect on these.

The NIME conferences accept different types of works such as papers (short and full), demos, music performances, installations, and workshops. The diversity of accepted submission types has also increased over the years. Each type of work is generally accompanied by a text document submitted by the authors, but there is a noticeable inconsistency in which types have been included in the annual proceedings. Figure 2 shows two dips in the average number of references per paper (cyan), suggesting that performance-related, or installation-related documents were included in the 2002 Dublin and 2009 Pittsburgh proceedings, as these typically contain few or no references. This trend is somewhat visible in the number of papers found in Semantic Scholar with complete (i.e. non-zero) reference data, as shown in Figure 1 (in blue). For some editions of the conference, non-paper works have been archived separately (with limited visibility), while, unfortunately, it appears that non-paper works were not systematically archived in some other cases. For the purpose of this analysis, filtering out non-paper works from the NIME paper proceedings BibTeX file is not possible because entries do not include information on the submission type, although an educated guess can be made in most cases.

In 2021 and 2022, NIME proceedings were published on PubPub[4], while earlier proceedings were published on Zenodo[5]. PubPub, a relatively new platform, allows media-rich publications and enables readers to comment on articles. The main limitation for this study is the absence of an archived, indexable PDF version of the paper that Semantic Scholar needs to analyze the text and extract the list of references, as shown by the data in Figure 1. A PDF version of the paper can be dynamically generated upon user request. The indexing of works published on PubPub poses problems for other search engines, such as Google Scholar, as these works are often duplicated. For example, Google Scholar struggles to merge the PubPub publication with the version of the PDF uploaded on institutional archives by authors, which could potentially inflate citation numbers. There are additional limitations with PubPub. While these are less relevant to this study, they do significantly impact the data generated by the NIME PA. For instance, author affiliations or emails are frequently missing, likely due to limitations with web rendering and perhaps a lack of clear instructions in the template. In a similar vein, the NIME PA converts papers from PDF to XML, but XML versions of papers are natively available in PubPub, which is an advantage. The PubPub URL of papers can be computed from the data in the NIME BibTeX file, but due to frequent changes to the PubPub front end, the method for extracting the XML download URL must be continuously updated. Additionally, it appears that PubPub has recently started blocking automated web scrapers.

We recommend addressing the discrepancy seen in the 2021 and 2022 proceedings archiving process, especially if PubPub is no longer being considered as a platform for publishing future works. For consistency with previous NIME proceedings, LaTeX sources could be downloaded from PubPub and traditional double-column PDFs could be generated. These PDFs could then be archived on the NIME website or on Zenodo. Any missing author information should ideally be retrieved from the conference management system, or by reaching out directly to the authors.

Scholars publish work at conferences to enhance the visibility of their research. The publication of proceedings extends this visibility beyond the physical venue of the conference itself. It is essential that the publication of proceedings in indexable repositories, along with the generation of associated ISBN or ISSN numbers and the assignment of DOIs for papers, occurs promptly by the end of the conference. Unfortunately, there have been delays that sometimes postpone the publication by nearly a year. Such procrastination not only risks compromising the indexing of NIME's proceedings—owing to the potential appearance of unofficial parallel versions uploaded by authors—but also diminishes the academic visibility of the works during the crucial first year after publication. Figure 12 illustrates that a NIME paper is more likely to be cited shortly after it becomes available, demonstrating the importance of timely proceedings for academic recognition.

## 5.2 Methodology limitations

Our computational analysis accounts for exceptions in the NIME proceedings archive and Semantic Scholar database that we have identified and reported, though they might not have been corrected. It is possible that some exceptions remain undetected or have not been adequately addressed.

We must acknowledge that the Semantic Scholar database may not be entirely complete or accurate. The natural language processing and machine learning techniques employed to extract information from PDFs of NIME indexed papers are not flawless. However, according to a recent study, Semantic Scholar's accuracy rate is estimated at 98.88% [20].

As mentioned earlier, Semantic Scholar extracts data primarily from PDFs of scholarly works that are freely accessible and not behind a paywall. However, PDFs shared on social networking sites for scientists and researchers are not included. Furthermore, the extraction of correct data from malformed PDFs is unlikely, and the NIME PA indicates that at least 10 papers may be affected by such issues.

Semantic Scholar's extraction of paper metadata, such as publication venue and publication type, is generally more reliable for journal articles. This is due to their consistent formatting and high quality, largely because they are not produced by the authors themselves. As a result, the accuracy of metadata for conference publications or older journal articles may be somewhat limited.

The 'field of study' tag is automatically inferred by Semantic Scholar, though the specifics regarding the accuracy of this model and the details of the dataset used for training are unknown. Furthermore, we cannot dismiss the possibility of biases or under-representation of certain disciplines in the training dataset.

Data on references and citations are retrieved and processed separately. Therefore, the number of references in NIME papers that are part of the NIME proceedings, and the number of citations to NIME papers from works in the NIME proceedings, are calculated independently. These numbers should technically be identical, but we find a minor discrepancy, with 5519 for the former and 5526 for the latter. However, this difference is slight, with a mismatch of just 0.13%, suggesting overall accuracy in Semantic Scholar's data and in our computational analysis.

## 5.3 Future Work

In future work, we aim to further develop the software used to retrieve publication-related data from existing databases and compute metrics related to references and citations. Specifically, we plan to introduce an easy-to-use text-based query mechanism for generating metrics and trends specific to a paper, field, author, or venue. While these might not be of broad interest, they could prove valuable to individual NIME researchers. Currently, this task is achievable but it requires navigation and manipulation of a collection of spreadsheets available online[1]. Additionally, introducing dynamic visualizations of selected data through the web could further facilitate the process of analyzing and interpreting NIME scholarly data. Furthermore, identical metrics can be computed for other publication venues in technologies for sonic art and music, allowing for direct comparison.

# 6. ACKNOWLEDGMENTS

I would like to thank Marcelo Wanderley for his comments on a preliminary draft of this work, and Ziyue Piao for compiling and sharing an early version of the NIME 2023 BibTeX file.

# 7. ETHICAL STANDARDS

This work has been conducted in accordance with all ethical and data protection guidelines set by the University of Oslo. Our analysis relies exclusively on secondary data available in public databases. While we strive for accuracy, we cannot guarantee the absolute correctness of the figures presented in this paper due to the potential for errors or inaccuracies within the databases we utilized. Specifically, the data retrieved from Semantic Scholar were extracted using machine learning models whose details have not been disclosed. Consequently, we cannot exclude the possibility of biases inherent in those models arising from the particular datasets used for their training. We affirm that we have no affiliation or conflict of interest with Semantic Scholar or its parent organization, the Allen Institute for AI. The computational analysis detailed in this paper is built upon open-source projects. We have greatly benefited from these projects and, in the spirit of contributing back, we are making our source code openly available to the public. The grammar and spelling checks were performed using the University of Oslo's GPT-3.5 Turbo service.

[4] https://nime.pubpub.org/

[5] https://zenodo.org/communities/nime_conference/